\documentclass[pre,aps,twocolumn,superscriptaddress]{revtex4-2}

\usepackage[pdftex]{graphicx}
\usepackage{amsbsy,amssymb,amsmath,bm,mathtools}

\usepackage{xspace}
\usepackage{bm}
\usepackage{color}
 
\usepackage{url}
\usepackage[section]{placeins}
\usepackage[colorlinks=true,linkcolor=blue,citecolor=blue]{hyperref}

\begin{document}


\title{Coarsening dynamics of fingerprint labyrinthine patterns: \\ Machine learning assisted characterization}

\author{Supriyo Ghosh}
\affiliation{Department of Physics, University of Virginia, Charlottesville, Virginia 22904, USA}
\affiliation{Present address: Department of Chemistry, University of Chicago, Chicago, IL 60637, USA}

\author{Vinicius Yu Okubo}
\affiliation{Dept. Electronic Systems Engineering, Polytechnic School, University of São Paulo, Brazil}

\author{Kotaro Shimizu}
\affiliation{Department of Applied Physics, The University of Tokyo, Tokyo 113-8656, Japan}

\author{ B. S. Shivaram}
\affiliation{Department of Physics, University of Virginia, Charlottesville, Virginia 22904, USA}

\author{Hae Yong Kim}
\affiliation{Dept. Electronic Systems Engineering, Polytechnic School, University of São Paulo, Brazil}

\author{Gia-Wei Chern}
\affiliation{Department of Physics, University of Virginia, Charlottesville, Virginia 22904, USA}

\date{\today}

\begin{abstract}
Fingerprint labyrinthine patterns exhibit a level of structural complexity beyond simple stripe phases, combining local stripe order with a dense network of point-like defects. Unlike symmetry-breaking phases, where coarsening proceeds via diffusive defect annihilation, or conventional stripe phases, where curvature-driven motion of extended grain boundaries dominates, the coarsening of fingerprint labyrinths is governed primarily by localized junction and terminal defects. Using the Turing-Swift-Hohenberg equation, we study the non-equilibrium relaxation of fingerprint labyrinthine patterns following a quench. To go beyond conventional Fourier-based diagnostics, we employ a template-matching convolutional neural network (TM-CNN) to identify and track junctions and terminals directly in real space, enabling a quantitative characterization of defect statistics and spatial correlations. We show that, although these point-like defects drive coarsening, their motion is strongly constrained by the surrounding stripe geometry, leading to slow, non-diffusive dynamics that are qualitatively distinct from both conventional phase ordering and stripe coarsening. Together, these results establish defect-mediated dynamics as the central organizing principle of fingerprint labyrinthine coarsening and demonstrate the effectiveness of machine-learning–assisted approaches for complex pattern-forming systems.
\end{abstract}

\maketitle

\section{Introduction}

In conventional symmetry-breaking phases, such as those described by Ising- or XY-type models, a quench from a disordered state produces domains characterized by a well-defined order parameter. The ensuing phase-ordering dynamics is marked by the growth of a characteristic domain size $L(t)$, which often follows an algebraic law $L(t) \sim t^\alpha$. The growth exponent $\alpha$ depends on basic properties of the system, including the symmetry of the ordered phase, whether the order parameter is conserved, and the nature of the relevant topological defects~\cite{bray1994,onuki2002,puri2009}. A unifying view is that phase ordering is driven by the motion, interaction, and annihilation of these defects---such as domain walls, vortices, or disclinations---which set the dominant length scale and lead to universal scaling behavior largely independent of microscopic details.

Pattern-forming systems differ qualitatively from this standard scenario. Instead of relaxing toward a uniform ordered state characterized by a homogeneous order parameter, they develop spatially modulated structures with a characteristic wavelength selected by a finite-wavenumber instability~\cite{cross2009,pisman2006,rabinovich2000,walgraef2012}. Prototypical examples include the Turing-Swift-Hohenberg (TSH)  class of models~\cite{swift1977, hohenberg1993}, which generate stripes, spots, and a wide range of more complex morphologies depending on parameters. Labyrinthine patterns, in particular, are disordered yet highly structured: they display strong local periodic order at a single wave number, typically manifested as a ring in the structure factor, but lack global phase or orientational coherence~\cite{seul1992c,elder1992,leberre2002,echeverria2020,tlidi2024}. Despite these differences, the organizing principle that ordering dynamics is governed by defect motion remains useful, with defects now embedded in and constrained by an underlying patterned background that fundamentally alters their dynamics.

This perspective is most clearly illustrated in the simplest pattern-forming case of stripe patterns, where coarsening is controlled by the motion of extended grain boundaries separating domains of different stripe orientation. Unlike curvature-driven domain-wall motion in Ising-type systems, where the interface velocity is set by the curvature of the wall itself, grain-boundary motion in stripe patterns is driven by elastic distortions of the surrounding pattern. Grain boundaries advance by reducing elastic energy stored in curved stripes, replacing them with straighter, lower-energy rolls. As a result, the driving force is determined by the curvature of stripes ahead of the boundary rather than by the boundary geometry itself. This mechanism leads to a distinct coarsening law, with a characteristic length scale growing as $L \sim t^{1/3}$~\cite{cross1995,hou1997,boyer2001,boyer2002, boyer2004, mazenko2006, igor2012}.

Labyrinthine patterns can be viewed as the next level of complexity beyond simple stripe states. While they retain a locally well-defined wavelength selected by the finite-wavenumber instability, they consist of a dense, interconnected network of curved stripe segments, junctions, and defects, and lack global orientational order. Such labyrinthine structures are ubiquitous across a wide range of non-equilibrium systems, ranging from chemical reaction–diffusion media and nonlinear optical cavities to magnetic films, liquid crystals, and biological tissues~\cite{lee1993,barrio1999,oswald2000,hardenberg2001,epstein2005,yochelis2008,liu2014}. Among the various labyrinthine morphologies, fingerprint-type labyrinths provide a natural and minimal extension of stripe patterns. These states arise when stripe coarsening is interrupted by the formation of bound defect complexes and extended amplitude grain boundaries, producing large domains of gently curved stripes arranged in fingerprint-like textures. In this sense, fingerprint labyrinths inherit key features of stripe systems such as elastic interactions and grain-boundary-mediated dynamics while introducing a controlled increase in structural and topological complexity.

In this paper, we present a comprehensive study of the coarsening dynamics of fingerprint labyrinthine patterns emerging from the TSH equation. To quantitatively characterize the ordering process, we combine conventional Fourier-based diagnostics such as structure factors and correlation lengths, with modern machine-learning techniques designed to extract emergent length scales and defect structures directly from spatial data. This hybrid approach enables us to assess the extent to which concepts from conventional phase-ordering theory can be generalized to labyrinthine patterns, and to identify dynamical features that are unique to this richer class of pattern-forming systems.

The remainder of this paper is organized as follows. Section~\ref{sec:TSH-review} introduces the TSH equation and the numerical methods used in this work. In Sec.~\ref{sec:fourier}, we characterize the coarsening dynamics using conventional global Fourier-based measures. Section~\ref{sec:TM-CNN} outlines a template-matching convolutional neural network (TM-CNN) algorithm to detect  point-like defects, included for completeness. In Sec.~\ref{sec:coarsening}, we analyze the coarsening dynamics based on the time evolution of these defects, while Sec.~\ref{sec:pair-dist} presents a pair-correlation analysis revealing their spatial organization. Section~\ref{sec:summary} concludes with a summary and outlook.

\section{Labyrinthine patterns from Turing-Swift-Hohenberg equation}

\label{sec:TSH-review}

The Turing-Swift-Hohenberg equation was originally introduced to study the role of thermal fluctuations near the onset of convective instabilities in Rayleigh--Benard systems~\cite{swift1977, hohenberg1993, dawes2016}. Although its initial formulation was motivated by stochastic effects close to a critical point, the deterministic TSH equation has since become a paradigmatic model for pattern formation in a wide range of physical, chemical and biological contexts. In its simplest deterministic form, the equation governs the relaxational dynamics of a scalar order parameter field $\psi(x,y,t)$ and reads
\begin{align}
\label{eq:SH}
	\frac{\partial \psi}{\partial t}	= \epsilon \psi - \psi^3 - \nu \nabla^2 \psi - \nabla^4 \psi ,
\end{align}
where $\epsilon$ is a bifurcation parameter controlling the distance from the pattern-forming threshold, and $\nu$ sets the relative importance of diffusive ($\nu<0$) or antidiffusive ($\nu>0$) contributions. The equation describes gradient-flow dynamics associated with a Lyapunov functional, so that the system evolves toward states of decreasing free energy. The competition between the linear instability $\epsilon\psi$, nonlinear saturation $-\psi^3$, and higher-order spatial derivatives leads to the spontaneous emergence of spatially modulated patterns with a characteristic wavelength.

The TSH equation~\eqref{eq:SH} is isotropic and invariant under spatial inversion $\psi\to-\psi$. Its trivial homogeneous solution $\psi=0$ is stable only within a limited region of parameter space. To analyze the onset of pattern formation, one considers small perturbations about the homogeneous state, $\psi=\delta\psi$ with $\delta\psi\ll1$, and substitutes a plane-wave ansatz $\delta\psi\sim e^{i\mathbf{k}\cdot\mathbf{r}-\omega t}$ into the linearized equation. This yields the dispersion relation $\omega(k)=\epsilon+\nu k^2-k^4$. The uniform state becomes unstable when $\omega(k)>0$ for a finite band of wave numbers. The fastest-growing mode occurs at the critical wave number $k_c=\sqrt{\nu/2}$, implying that for $\epsilon>-\nu^2/4$ the homogeneous state is unstable and the system evolves toward spatially periodic structures with a preferred wavelength $2\pi/k_c$. Near threshold, this instability produces stripe patterns, while farther from onset nonlinear interactions and defect dynamics generate increasingly complex morphologies.

\begin{figure*}
\centering
\includegraphics[width=1.99\columnwidth]{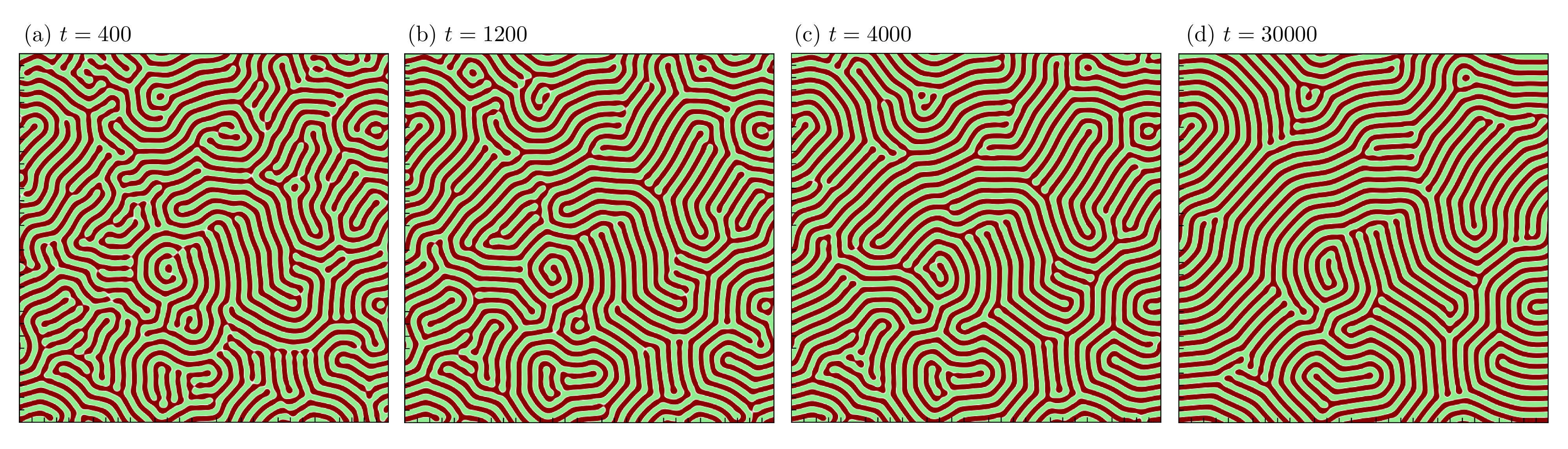}
\caption{Coarsening dynamics of the fingerprint labyrinthine phase of the TSH equation at increasing times. Coarsening is dominated by stripe breaking and reconnection events, corresponding to the annihilation of junctions and stripe terminals, while the motion of extended grain boundaries is strongly suppressed. The evolution is governed by local topological rearrangements within the stripe network rather than curvature-driven domain growth.}
\label{fig:coarsening}
\end{figure*}

Beyond serving as a minimal model for stripe formation, the TSH equation supports a rich hierarchy of spatially ordered and disordered steady states as the bifurcation parameter increases \cite{echeverria2020, katsuya2006}. In the parameter regime closest to onset, the system relaxes toward a pure stripe phase, characterized by long-range orientational order and coarsening governed by the motion of extended grain boundaries \cite{boyer2004, mazenko2006}. At larger driving, stripe coarsening is interrupted by defect binding and pinning, giving rise to fingerprint-type labyrinthine patterns \cite{tlidi2024}. These states consist of extended domains of gently curved stripes separated by amplitude grain boundaries; they retain a single dominant wavelength and strong local periodic order, yet lack global orientational coherence. 

As such, fingerprint labyrinths represent the simplest and most natural extension of stripe patterns, preserving the underlying stripe-based elastic mechanisms while introducing a controlled increase in topological and structural complexity. Upon further increase of the control parameter, the system enters more strongly disordered labyrinthine regimes, including glassy labyrinths dominated by frozen local defects and scurfy labyrinths characterized by the proliferation of spotlike structures embedded within the patterned background. In this work, we focus on the fingerprint labyrinthine phase as an intermediate and physically transparent regime, ideally suited for probing how conventional phase-ordering mechanisms are modified, frustrated, or arrested as pattern complexity increases.

Numerical integration of the TSH equation is a well-studied problem that requires careful treatment due to the coexistence of nonlinear terms and high-order spatial derivatives. In particular, the biharmonic term $\nabla^4\psi$ introduces stiffness that imposes restrictive stability limits on explicit time-stepping schemes. Fully implicit methods provide unconditional stability but involve the solution of nonlinear systems at each time step, which can become computationally demanding for large system sizes or long-time simulations. In this work, we adopt a standard semi-implicit time-integration scheme~\cite{Lee2017}, in which the linear terms are treated implicitly, while the nonlinear term $\psi^3$ is linearized with respect to the current solution, offering an effective compromise between numerical stability and computational efficiency.

Specifically, denoting the solution at time step $n$ by $\psi^{n}$, the nonlinear term is approximated using a first-order Taylor expansion, $(\psi^{n+1})^3 \simeq 3(\psi^{n})^2 \psi^{n+1} - 2(\psi^{n})^3$, which renders the update equation linear in $\psi^{n+1}$. The resulting linear system is solved iteratively at each time step until convergence is achieved. This semi-implicit iterative approach permits substantially larger time steps than explicit schemes, while avoiding the computational cost of fully nonlinear solvers. It is particularly well suited for capturing the long-time evolution of the system, including slow coarsening dynamics, defect interactions, and the emergence of metastable labyrinthine configurations. By combining implicit treatment of stiff linear operators with controlled linearization of nonlinear terms, this method provides a robust and efficient framework for simulating the deterministic dynamics of the TSH equation.

Fig.~\ref{fig:coarsening} illustrates the coarsening dynamics of the fingerprint labyrinthine phase of the TSH equation over several decades in time. A defining feature of the evolution is the persistence of a well-defined local stripe wavelength, even as the global pattern gradually simplifies. Unlike conventional stripe coarsening near onset—where domains of distinct orientation grow through the steady motion of extended grain boundaries—large-scale reorganization of stripe orientations is strongly suppressed in this regime. Instead, the pattern retains its labyrinthine character over long time, signaling a qualitatively different coarsening mechanism.

Coarsening proceeds primarily through localized topological rearrangements of the stripe network. The dominant processes involve the breaking and reconnection of stripe segments, which can be naturally interpreted as the annihilation of point-like defects such as stripe junctions and stripe terminals. These events locally reduce the defect density and simplify network connectivity without requiring coherent motion of extended interfaces. As a consequence, the characteristic length scale increases only slowly, reflecting dynamics controlled by rare spatially localized defect interactions rather than by continuous curvature-driven relaxation.

This behavior can be traced to the effective suppression of grain-boundary motion, a hallmark of the fingerprint labyrinthine phase. Extended amplitude grain boundaries, which govern stripe coarsening near threshold, appear pinned or dynamically arrested, preventing the system from evolving toward a globally ordered stripe state on accessible time scales. The dynamics instead proceeds through a sequence of metastable configurations connected by discrete topological defect events, leading to slow, glassy-like relaxation despite the absence of quenched disorder. Together, these observations place the fingerprint labyrinthine phase in an intermediate regime between simple stripe patterns and strongly disordered labyrinthine states, where the stripe-based elastic framework remains relevant, but coarsening is governed by point-defect annihilation rather than smooth interface motion.

\section{Global Fourier analysis }

\label{sec:fourier}

\begin{figure*}
\centering
\includegraphics[width=1.99\columnwidth]{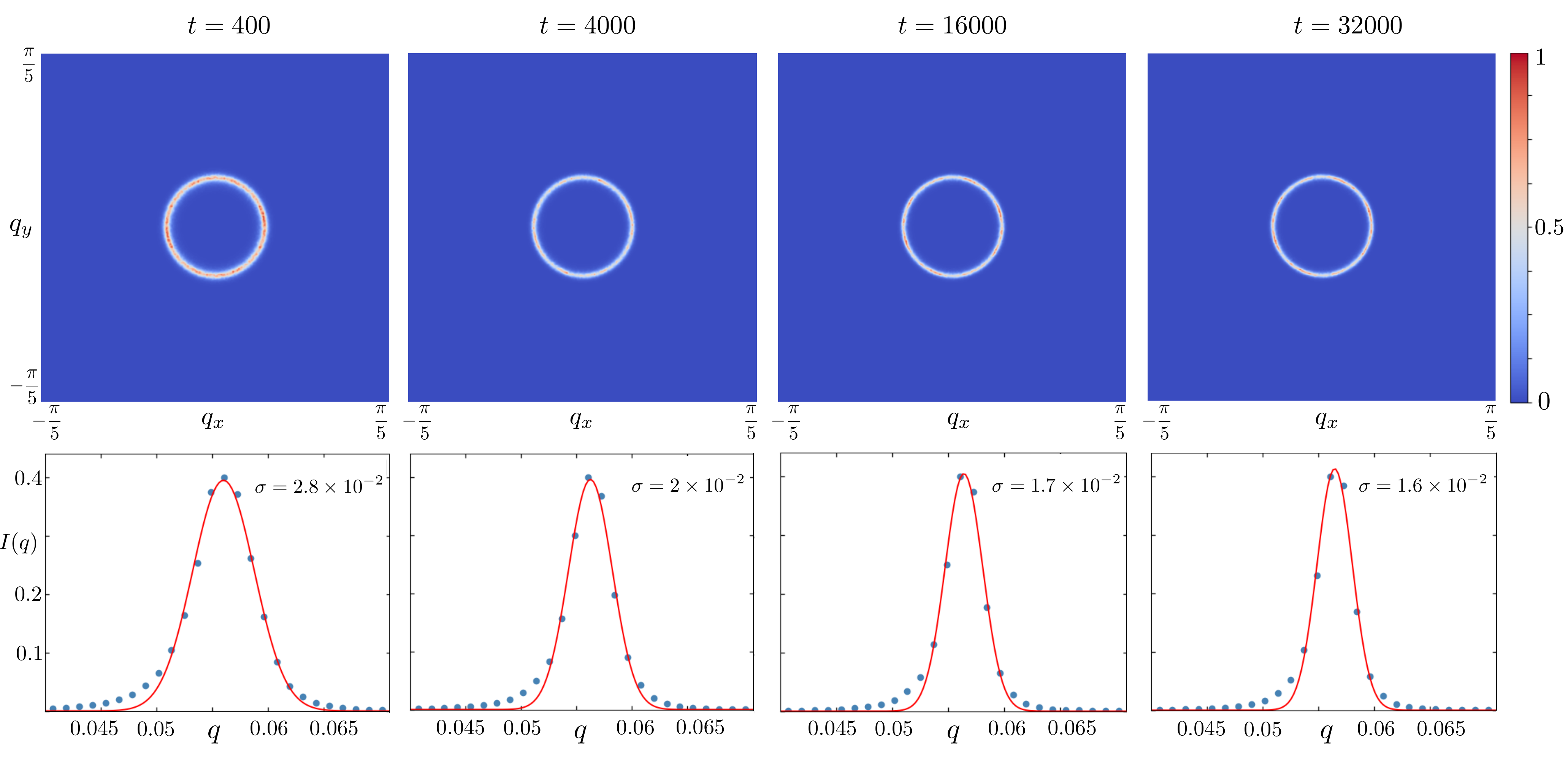}
\caption{Time evolution of the global structure factor for the fingerprint labyrinthine phase obtained from simulations of the TSH equation. The top panels show the two-dimensional structure factor $S(\mathbf{q})$ at representative times, displaying a ring-like distribution centered at a well-defined wave number. The bottom panels show the corresponding angularly averaged structure factor $I(q) \sim \exp[-(q - q_c)^2/2\sigma^2]$. Solid lines indicate Gaussian fits to $I(q)$, and $\sigma$ denotes the standard deviation of the fit, which characterizes the radial width of the ring.
}
\label{fig:fourier}
\end{figure*}

Fourier-based analysis has long served as a standard and powerful framework for characterizing spatial order in pattern-forming systems governed by finite-wavenumber instabilities, such as Rayleigh–Bénard convection, reaction–diffusion models, and the TSH equation~\cite{cross1995,boyer2001,boyer2002,ouyang1991,christensen1998}. In these systems, the emergence of structure is naturally associated with the selection of a characteristic wave number, making the global Fourier spectrum a primary diagnostic to identify ordered phases, track coarsening dynamics, and distinguish between stripe, hexagonal, and disordered states~\cite{cross2009,pisman2006,rabinovich2000,walgraef2012}. In stripe-forming regimes, global Fourier analysis has been particularly successful in linking real-space coarsening dynamics to spectral evolution. Near the onset of pattern formation, long-range orientational order manifests as sharp peaks in the Fourier spectrum at well-defined wave numbers and orientations. As domains grow and grain boundaries annihilate, these peaks progressively narrow in time, enabling the extraction of characteristic length scales and dynamic scaling laws~\cite{cross1995,boyer2001,boyer2002}. This spectral viewpoint underpins much of the conventional understanding of stripe coarsening and provides a natural reference point for analyzing more complex pattern-forming regimes.

The global Fourier transform is defined by computing the spatial Fourier spectrum of the entire order-parameter field,\begin{equation}
	\tilde{\psi}(\mathbf{q}) =	\frac{1}{L^2}\int d^2 r \, \psi(\mathbf{r})\, e^{-i\mathbf{q}\cdot\mathbf{r}},
\end{equation}
with the corresponding structure factor given by
\begin{equation}
	S(\mathbf{q}) = |\tilde{\psi}(\mathbf{q})|^2 .
\end{equation}
Because this transform averages over the full system, it captures global ordering tendencies while discarding information about spatial heterogeneity. For stripe phases, the resulting spectrum exhibits pronounced peaks at wave vectors $\pm\mathbf{q}_c$, reflecting both wavelength selection and global orientational order. In contrast, labyrinthine patterns produce a qualitatively different spectral signature: rather than discrete angular peaks, the spectrum displays a ring-like structure centered at a single dominant wave number. The persistence of a sharp radial peak indicates robust local wavelength selection, while the absence of angular structure signals the loss of global orientational coherence. This characteristic ``powder-ring'' spectrum has become a defining hallmark of labyrinthine states~\cite{leberre2002,echeverria2020}.

The structure-factor evolution in Fig.~\ref{fig:fourier} illustrates these features for the fingerprint labyrinthine phase. At all times, the two-dimensional structure factor $S(\mathbf{q})$ displays a pronounced ring at a well-defined wave number, with no discrete orientational peaks. This ring reflects strong local periodic order coexisting with global orientational disorder, inherited from the stripe-forming instability. Its persistence shows that coarsening does not involve large-scale alignment of stripe domains, but instead preserves the labyrinthine character of the pattern.

During relaxation, the dominant change in $S(\mathbf{q})$ is a gradual sharpening of the ring: its radial width decreases while the radius remains essentially fixed. This indicates that the selected wavelength is conserved, whereas fluctuations around it are progressively suppressed. Such spectral narrowing is consistent with the slow elimination of defects and local distortions within the interconnected stripe network, leading to increased spatial coherence without the development of long-range orientational order~\cite{cross1995,boyer2001}.

\begin{figure}
\centering
\includegraphics[width=.99\columnwidth]{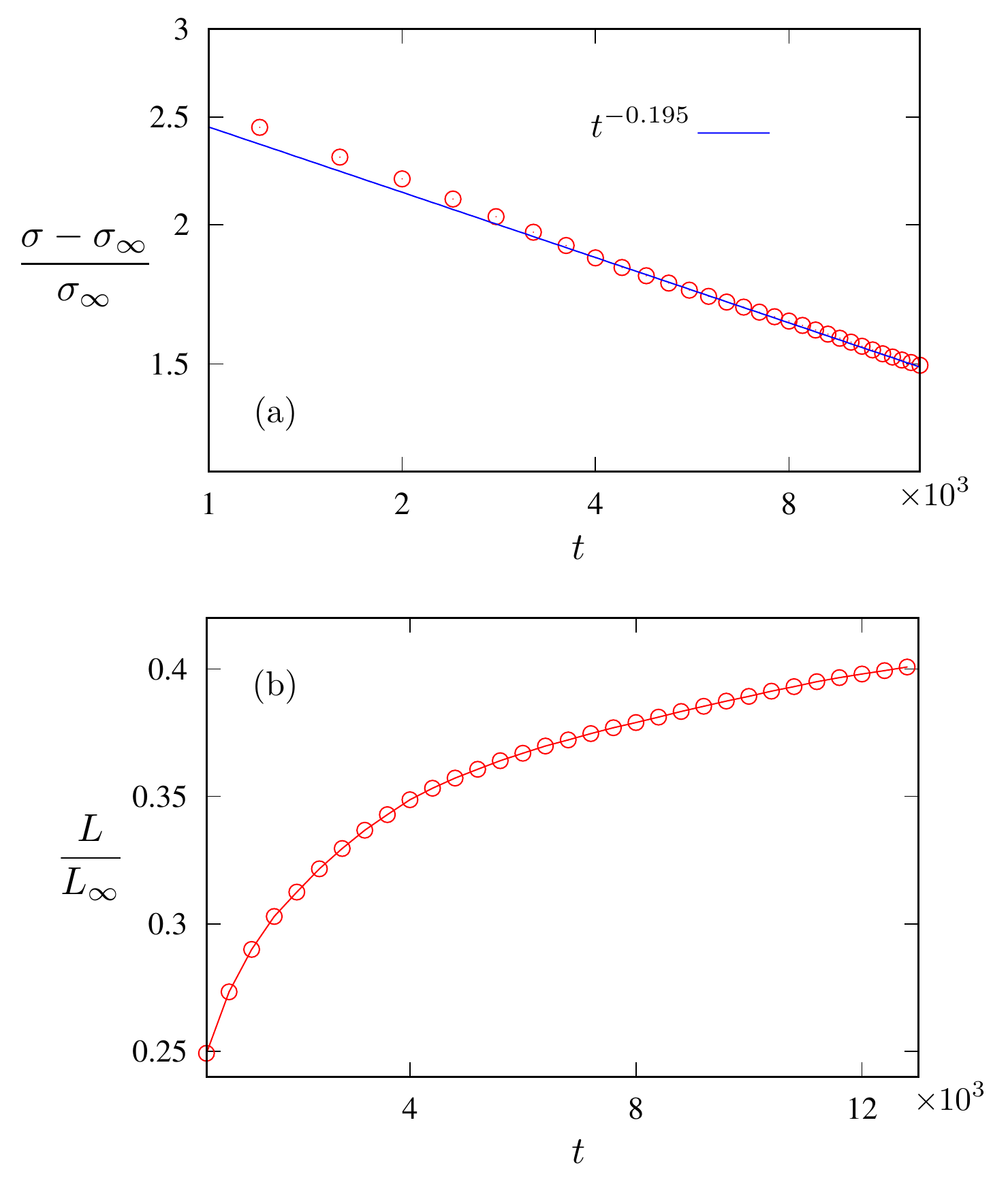}
\caption{
(a) Time evolution of the ring width $\sigma$ in the angular-averaged structure factor of the fingerprint labyrinthine pattern, shown as the normalized deviation $(\sigma-\sigma_\infty)/\sigma_\infty$, where $\sigma_\infty$ is the long-time limit. The relaxation is well described by a power-law decay (solid line). (b) Corresponding evolution of the correlation length $L \sim 1/\sigma$, normalized by its asymptotic value $L_\infty$, showing slow growth associated with the narrowing of the structure-factor ring. 
}
\label{fig:clen}
\end{figure}

To quantify this evolution, we analyze the angularly averaged structure factor $I(q)$, obtained by averaging $S(\mathbf{q})$ over angles. The resulting spectra exhibit a single dominant peak, well described by a Gaussian over the relevant range of wave numbers (Fig.~\ref{fig:fourier}, bottom). From these fits, we extract the standard deviation $\sigma(t)$, which measures the radial width of the spectral ring. At very late times, the coarsening dynamics enters a frozen regime, signaled by the saturation of $\sigma(t)$ and indicating incomplete ordering. To account for this arrest, we include a residual width $\sigma_\infty$ and fit the data with
\begin{eqnarray}
	\label{eq:sigma-t}
	\sigma(t) = \sigma_\infty + A t^{-\alpha}, 
\end{eqnarray}
yielding $\alpha \simeq 0.195$ (Fig.~\ref{fig:clen}(a)). The excellent agreement demonstrates a slow power-law relaxation toward the frozen state, highlighting the role of long-lived defects or kinetic constraints intrinsic to the labyrinthine morphology.

Fig.~\ref{fig:clen}(b) shows the corresponding evolution of the characteristic correlation length $L$, defined through the inverse relation~\cite{morris1993,egolf1998,echeverria2020}
\begin{eqnarray}
	L(t) \sim 1/\sigma(t), 
\end{eqnarray}
which may be interpreted as the typical stripe or domain size in the pattern. Consistent with the slow decay of $\sigma(t)$, the correlation length exhibits a gradual growth over time and approaches its asymptotic value only algebraically. This behavior reflects the increasingly sluggish nature of the late-stage coarsening dynamics, where further domain growth is strongly hindered, and provides a complementary real-space perspective on the arrested coarsening inferred from the structure-factor analysis.

While the global Fourier analysis provides a clear quantitative description of the late-stage coarsening in terms of length scales and spectral narrowing, it offers limited insight into the microscopic mechanisms that control the relaxation process. In particular, our simulations indicate that the slow power-law approach to the arrested state is not dominated by grain-boundary motion or by large-scale reorientation of stripe domains. For these reasons, a local or windowed Fourier-transform analysis~\cite{egolf1998,echeverria2020} and local disorder measures~\cite{gunaratne1995,hu2004,hu2005}---often useful in conventional stripe-forming systems---does not provide additional meaningful information in the present labyrinthine regime.

Instead, direct inspection of the dynamics reveals that relaxation proceeds primarily through highly localized processes involving the breaking and reconnection of stripes. These events are mediated by point-like defects, such as junctions and stripe terminals, whose creation, annihilation, and rearrangement govern the long-time evolution of the pattern. Accurately identifying and tracking these defects is therefore essential for uncovering the microscopic pathways underlying the arrested coarsening. To this end, we employ the recently developed template-matching convolutional neural network (TM-CNN  \cite{okubo2024}), which provides an efficient and robust framework for detecting and classifying point-like defects.

\section{Defect detection with template matching and CNN}

\label{sec:TM-CNN}

\begin{figure}[b]
\centering
\includegraphics[width=0.9\columnwidth]{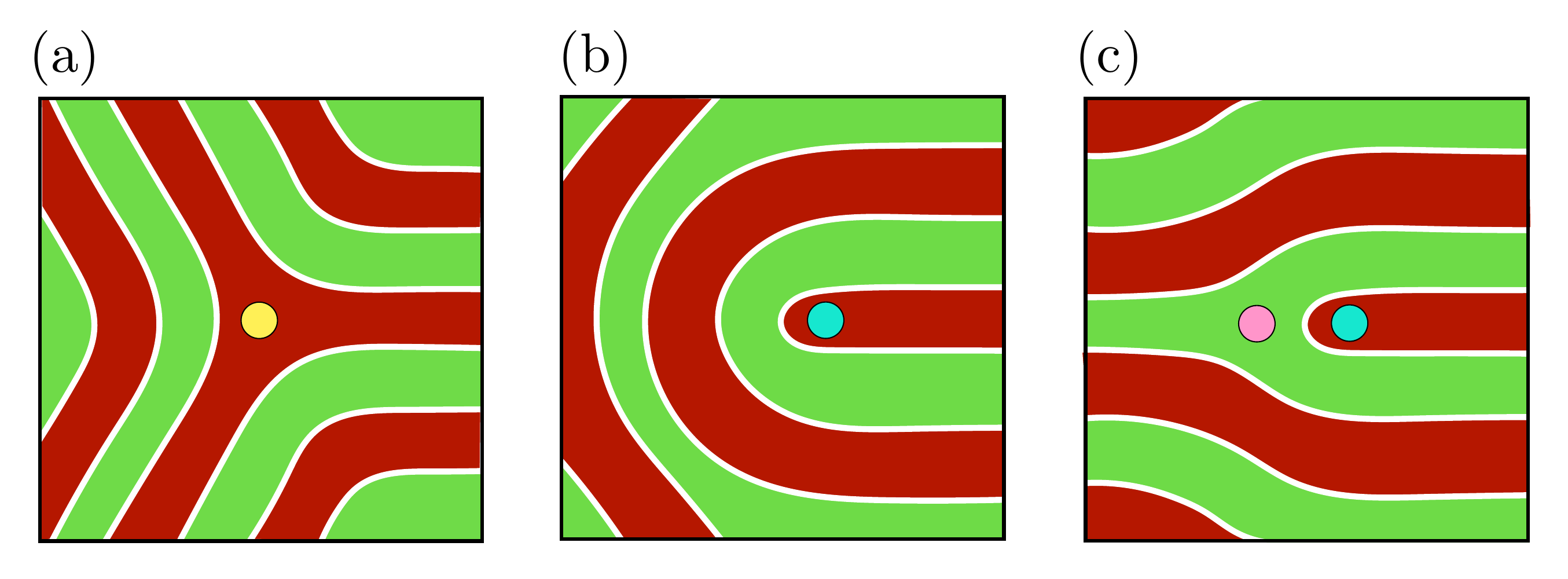}
\caption{Schematic of point-like defects in stripe patterns. (a) Junction defect (red-stripe viewpoint), corresponding to a $-1/2$ disclination in an effective orientational description. (b) Terminal defect, corresponding to a $+1/2$ disclination, with the sign set by the winding of the local stripe orientation. (c) Bound pair of terminal (blue-stripe) and junction (red-stripe) forming a composite defect with zero net disclination charge, behaving as an effective dislocation of the stripe pattern.}
\label{fig:defects}
\end{figure}

\begin{figure*}
\centering
\includegraphics[width=1.99\columnwidth]{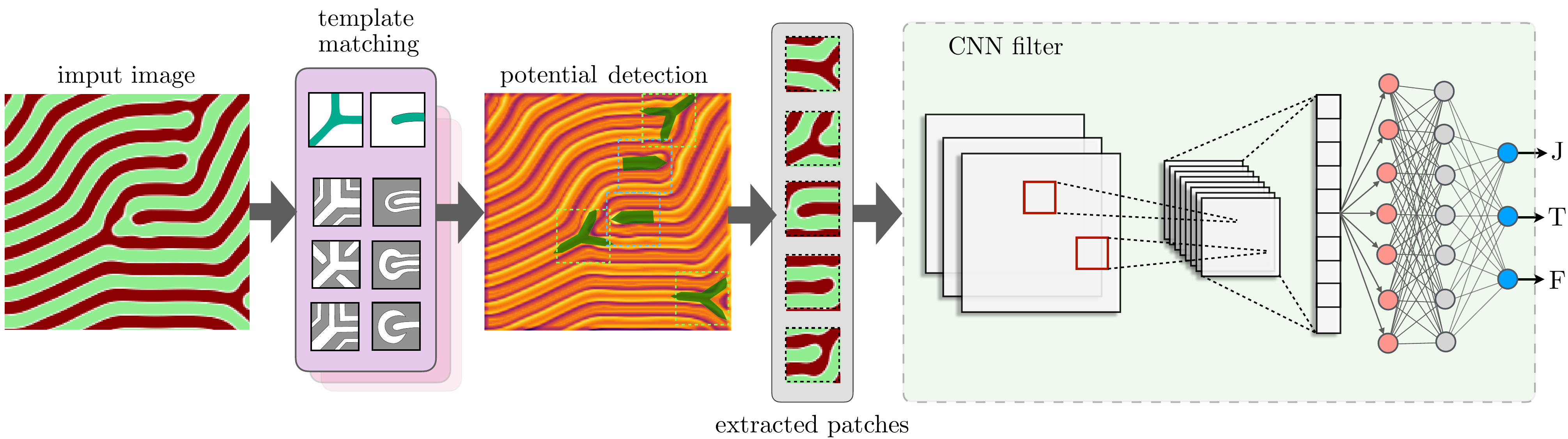}
\caption{Schematic illustration of the template-matching convolutional neural network (TM-CNN) workflow used to identify point-like defects in labyrinthine patterns. Starting from the input image, a set of predefined templates is used to perform template matching, generating correlation maps that highlight candidate defect locations. A low detection threshold is employed to ensure that all potential junctions and terminals are captured, followed by non-maximum suppression to remove duplicate detections. Image patches centered on these candidate locations are then extracted and passed to a convolutional neural network, which classifies each patch as a junction (J), a terminal (T), or a false detection (F). This two-stage procedure combines the sensitivity of template matching with the selectivity of a CNN, enabling efficient and accurate detection of point defects.}
\label{fig:TM-CNN}
\end{figure*}

The fingerprint labyrinthine pattern contains a small set of elementary point-like defects, most prominently junctions and terminals, which serve as the primary agents of local pattern rearrangement. Representative examples of a junction and a terminal are shown schematically in Fig.~\ref{fig:defects}(a) and (b), respectively. An important subtlety is that the classification of a given defect as a junction or a terminal depends on whether it is defined with respect to the red or green stripes. In the TSH description, these correspond to positive and negative values of the scalar field $\psi(\mathbf{r})$ and are related by a global $Z_2$ symmetry. Exchanging red and green stripes is equivalent to the transformation $\psi \rightarrow -\psi$, under which the roles of junctions and terminals are interchanged. For simplicity and without loss of generality, we adopt the red-stripe convention throughout this work, as illustrated in Fig.~\ref{fig:defects}(a) and (b).

These defects admit a natural topological interpretation in terms of the director field $\mathbf{n}(\mathbf{r}) = \boldsymbol{\nabla}\psi / |\boldsymbol{\nabla}\psi|$, which encodes the local stripe orientation~\cite{seul1992a,seul1992b,seul1992d}. Within this framework, the junction and terminal shown in Fig.~\ref{fig:defects}(a) and (b) correspond to disclinations of the director field with topological charges $-1/2$ and $+1/2$, respectively, reflecting the winding of the stripe orientation around the defect core. Of particular importance is the composite defect shown in Fig.~\ref{fig:defects}(c), formed by a red-stripe terminal bound to a green-stripe junction. This composite carries zero net disclination charge and instead corresponds to a dislocation, associated with a finite Burgers-vector--like mismatch of the stripe phase~\cite{mermin1979}. Such dislocations play a distinct role from isolated disclinations, mediating translational rearrangements while preserving the overall orientational topology of the labyrinthine state.

To detect junctions and terminals, we employ the TM-CNN framework \cite{okubo2024}, which presents two-stage approach that combines classical image-processing techniques with convolutional neural network classification. The overall workflow is schematically illustrated in Fig.~\ref{fig:TM-CNN}. We analyze grayscale fingerprint images of size $600 \times 600$ pixels, scaled to the $[0,255]$ intensity range. This resolution resolves the fine ridge structure of the patterns while maintaining manageable computational cost. The method is well suited to the present problem, as it efficiently handles a high density of small-scale features while substantially reducing the need for manual annotation.

In the first stage, candidate locations for junctions and terminals are identified using template matching. A set of small templates representing typical ridge bifurcations and terminations is scanned across the entire image, producing correlation maps that highlight regions resembling these defect motifs, as indicated in Fig.~\ref{fig:TM-CNN}. To minimize false negatives, a deliberately low detection threshold is applied at this stage, ensuring that essentially all true defects are captured, although at the cost of introducing a significant number of false positives. A non-maximum suppression procedure described in~\cite{okubo2024} is then applied to the correlation maps to remove redundant detections and retain only the most likely candidate points, yielding a compact and tractable set of potential defect locations.

In the second stage, these candidate points are passed to a convolutional neural network (CNN) to classify each detection as a true junction, a true terminal, or a spurious false detection. For each candidate, a small image patch centered on the detected location is extracted and used as input to the CNN, as schematically shown in Fig.~\ref{fig:TM-CNN}.
This two-stage design combines high sensitivity in the initial detection phase with high precision in the final classification, overcoming the limitations of either template matching or CNN-based detection when used in isolation.

A key practical advantage of TM-CNN lies in its efficient labeling strategy for model training. 
Rather than manually labeling every junction and terminal in densely populated fingerprint images a task that would be prohibitively time-consuming template matching provides an initial set of candidate detections that require only limited human correction. These semi-automatically curated annotations are then used to train the CNN classifier, yielding a robust detection model with minimal manual effort. A detailed description of the full TM-CNN algorithm, along with systematic benchmarking against alternative methods, is provided in Ref.~\cite{okubo2024}.

In certain situations, such as during the formation or annihilation of defects, junctions and terminals can become ambiguous and difficult to classify. In these regimes, TM-CNN may produce inconsistent labels, as junction and terminal classifications are performed independently and the local image features may lie near the decision boundary of the CNN. To enforce topological consistency,  the medial axis skeleton~\cite{blum1967} of the fingerprint patterns are extracted and used as an auxiliary constraint, with domains segmented by thresholding the midpoint of the images intensity range. Skeletons provide an unambiguous topological identification of junctions and terminals, but they are prone to spurious branches arising from small fluctuations in stripe geometry. These artifacts are removed by matching TM-CNN detections to the skeleton and trimming unmatched branches, yielding a consistent and physically meaningful defect identification, as described in Ref.~\cite{okubo2025}.

\section{Relaxation dynamics and stripe domain coarsening}

\label{sec:coarsening}

\begin{figure}
\centering
\includegraphics[width=.99\columnwidth]{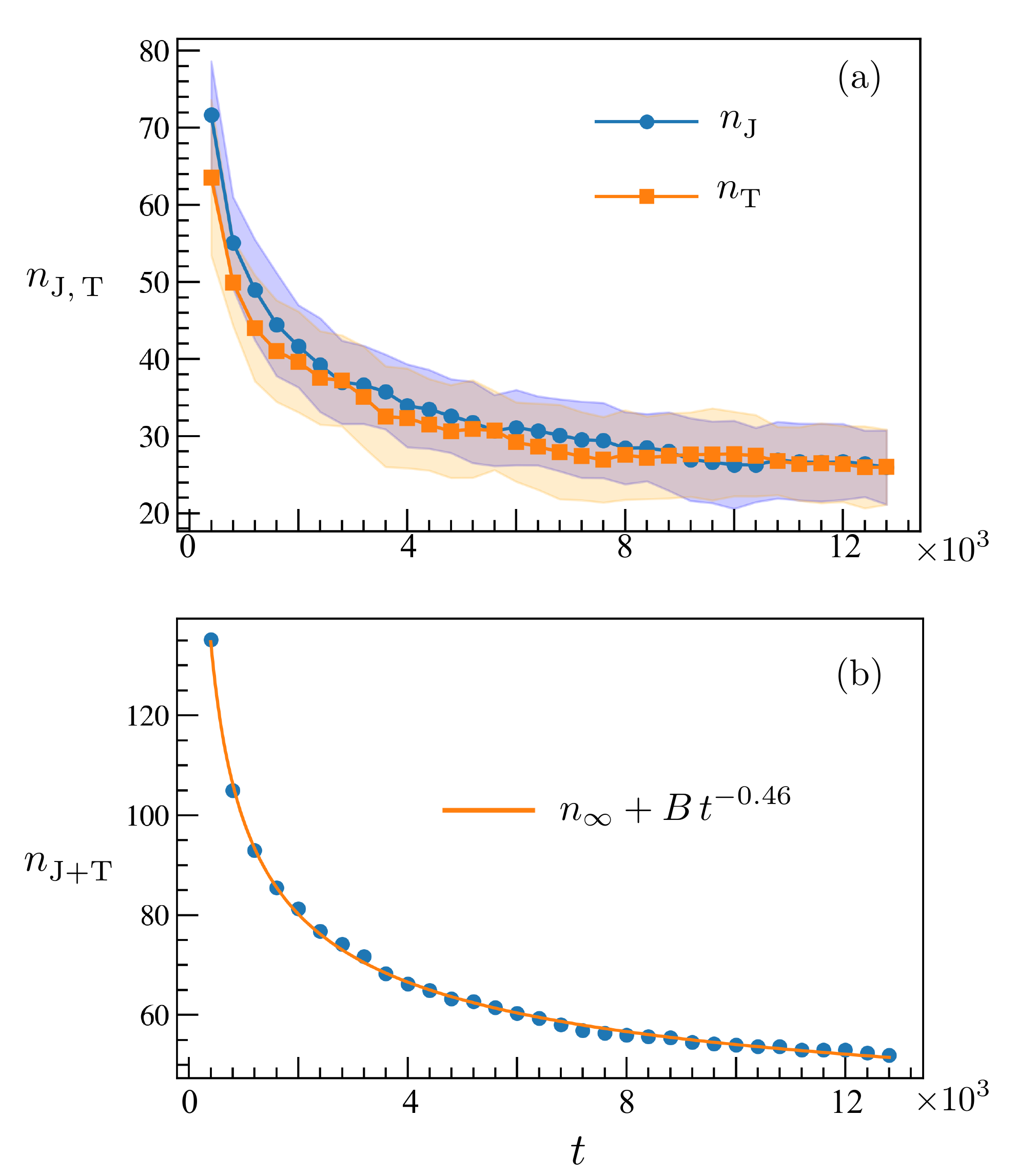}
\caption{Time evolution of point-like defects in the fingerprint labyrinthine pattern. (a) Number of junctions $n^{\,}_{\rm J}$ and terminals $n^{\,}_{\rm T}$ as a function of time, with shaded bands indicating fluctuations across independent realizations. Both defect types decay slowly and remain comparable in number throughout the evolution. (b) Total number of point-like defects $n^{\,}_{\rm J+T}=n^{\,}_{\rm J}+n^{\,}_{\rm T}$, showing a gradual decay toward a finite asymptotic value $n^{\,}_{\infty}$. The solid line indicates a power-law fit with exponent $\simeq 0.46$.}
\label{fig:n-vs-t}
\end{figure}

The evolution of point-like defects in the fingerprint labyrinthine pattern is summarized in Fig.~\ref{fig:n-vs-t}. Panel~(a) shows the time dependence of the number of junctions ($n^{\,}_{\rm J}$) and terminals ($n^{\,}_{\rm T}$), with shaded bands indicating fluctuations across independent realizations. At early times, both defect types are abundant, reflecting the highly disordered initial state. As coarsening proceeds, their numbers decrease rapidly due to frequent creation, annihilation, and recombination of junction-terminal pairs. Notably, $n^{\,}_{\rm J}$ and $n^{\,}_{\rm T}$ decay in close tandem throughout the evolution, reflecting the fact that the two defect types can be continuously transmuted into one another during the relaxation process, as discussed below.

At late times, however, the decay of defect populations slows dramatically and eventually arrests. This behavior mirrors the freezing observed in the structure-factor evolution and indicates that the labyrinthine pattern does not relax to a defect-free state. Instead, a finite density of long-lived defects survives, reflecting kinetic constraints and topological trapping intrinsic to the fingerprint morphology. To quantify this arrested relaxation, we analyze the total number of point-like defects, $n^{\,}_{\rm J+T}=n^{\,}_{\rm J}+n^{\,}_{\rm T}$, shown in Fig.~\ref{fig:n-vs-t}(b). The data are well described by a power-law approach to a nonzero asymptote,
\begin{eqnarray}
	n^{\,}_{\rm J+T}(t) = n_\infty + B t^{-\beta},
\end{eqnarray}
where $n_\infty$ represents the residual defect density in the frozen state. The fit yields an exponent $\beta \simeq 0.47$, indicating a slow algebraic relaxation toward arrest rather than exponential decay.

The observed power-law decay of the defect population might be expected to parallel the decay of the inverse correlation length, $\sigma \sim L^{-1}$, as given in Eq.~(\ref{eq:sigma-t}). In conventional symmetry-breaking systems, such as XY-type models, the characteristic correlation length is often set by the typical separation between topological defects, leading to the relation $L \sim n^{-1/d}$, where $d$ denotes the spatial dimension. If this reasoning were applicable here, it would imply a scaling relation between the corresponding exponents, $\alpha = \beta/2$, where $\alpha$ governs the decay of~$L^{-1}$ in Eq.~(\ref{eq:sigma-t}). 

However, the exponents extracted from our simulations clearly violate this relation. This discrepancy can be partly attributed to the contribution of grain-boundary motion, which is not directly tied to the dynamics of point-like defects, rather to the overall coarsening process of the fingerprint labyrinthine pattern. More fundamentally, as we demonstrate below, the relation $L \sim n^{-1/d}$ relies on the assumption that point defects are distributed uniformly in space, an assumption that is not satisfied in the present system.

A closer inspection of the simulations reveals that the dynamics of junction (J) and terminal (T) defects are highly constrained. This behavior contrasts sharply with the diffusive, random-walk like motion of topological defects typically observed in conventional symmetry-breaking phases. In the fingerprint labyrinthine pattern, defect motion is strongly coupled to the surrounding stripe texture and proceeds through localized, collective stripe rearrangements rather than through independent defect diffusion. Consequently, defect interactions follow specific, geometry-dependent pathways.

\begin{figure}
\centering
\includegraphics[width=1.0\columnwidth]{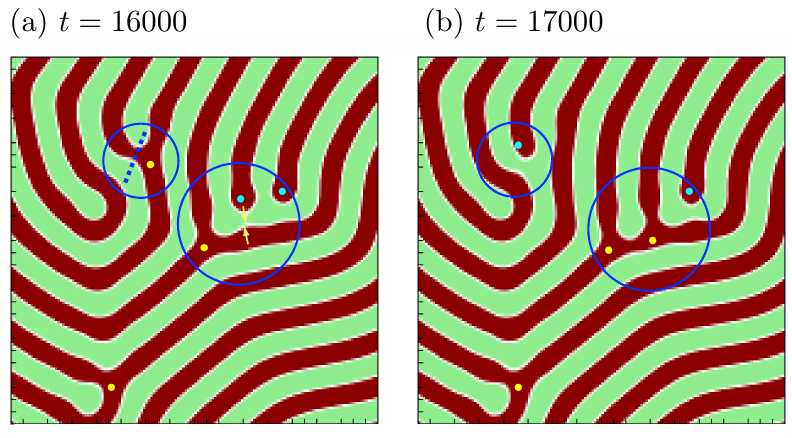}
\caption{Interconversion of junctions and terminals in a fingerprint labyrinthine pattern. Snapshots at (a) $t=16000$ and (b) $t=17000$ illustrate local stripe rearrangements that convert a junction into a terminal (left circle) and, conversely, a terminal into a junction (larger right circle). }
\label{fig:conv}
\end{figure}

Fig.~\ref{fig:conv} illustrates two representative defect-transmutation events that exemplify this constrained dynamics. In the left panel (smaller circle), a junction is converted into a terminal through breaking a red stripe, as indicated by the blue dashed line. This process alters the local stripe connectivity and changes the winding of the director field around the defect core, thereby transforming the topological character of the point-like defect without requiring long-range motion. The larger circled region on the right shows the inverse process, in which a terminal is converted into a junction through the formation and reconnection of a red stripe segment. These examples demonstrate that junctions and terminals are not topologically immutable objects but can be continuously transmuted into one another via local stripe breaking and reconnection. Such interconversions preserve the global topological charge while enabling local relaxation of elastic distortions.

\begin{figure}
\centering
\includegraphics[width=1.0\columnwidth]{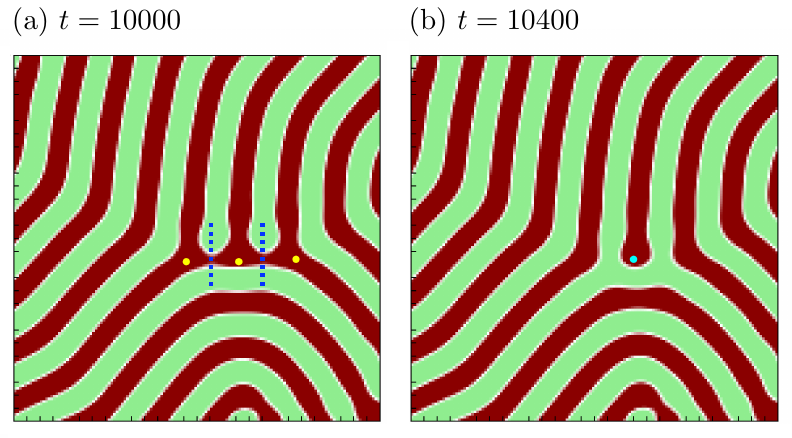}
\caption{Reduction of multiple junctions to a single terminal via stripe breaking. Snapshots at (a) $t=10000$ and (b) $t=10400$ show an event in which three junctions merge and are eliminated through the breaking of the red stripes, resulting in the formation of a single terminal. }
\label{fig:anni}
\end{figure}

Fig.~\ref{fig:anni} highlights a second class of defect interactions involving tightly bound clusters of junctions. In particular, we observe configurations in which three J-type defects, junctions, associated with a local winding of approximately $-\pi$ (corresponding to charge $-1/2$), aggregate within a small region. In the example shown, the cluster effectively combines two $-1/2$ charges and one $+1/2$ charge, resulting in a net charge of $-1/2$. Although such clusters are difficult to resolve visually due to their close proximity, they are dynamically significant. These triplets typically arise in regions of geometric frustration where competing stripe orientations meet and often serve as transient intermediates that subsequently collapse into a single terminal or participate in further annihilation events.

Taken together, our simulations indicate that nearly all defect motion and defect-number reduction in the fingerprint labyrinthine pattern proceeds through two elementary mechanisms: local junction-terminal transmutation and the reduction of multi-junction clusters (such as three-to-one events) that effectively annihilate defects. These processes, rather than free diffusion of isolated defects, constitute the dominant microscopic pathways responsible for the slow algebraic decay and eventual saturation of the total defect density $n^{\,}_{\rm J+T}$.


\section{Pair distribution function}

\label{sec:pair-dist}

\begin{figure}
\centering
\includegraphics[width=1.\columnwidth]{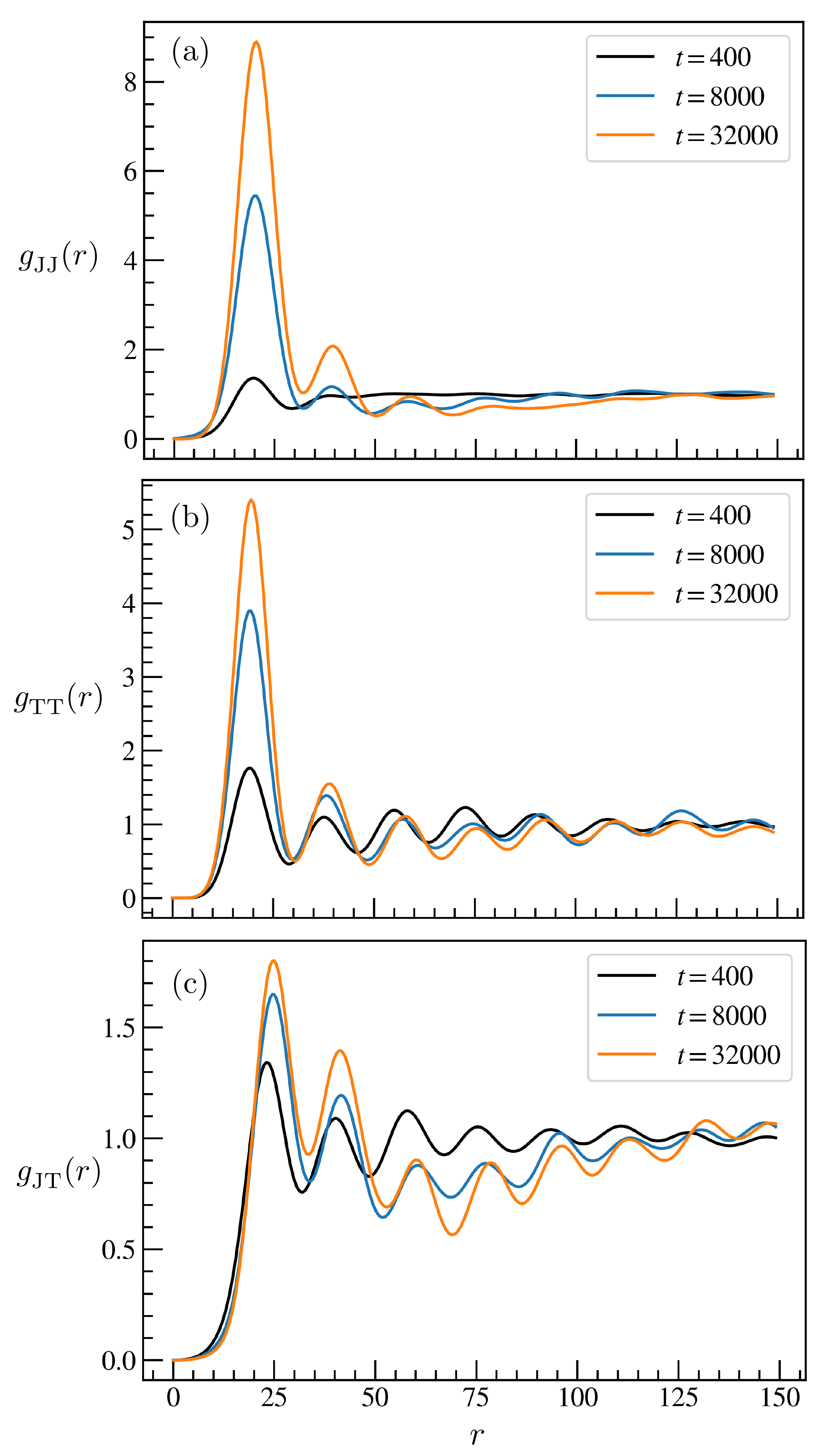}
\caption{Radial pair correlation functions for (a) junction-junction pairs, $g^{\,}_{\rm JJ}(r)$, (b) terminal-terminal pairs, $g^{\,}_{\rm TT}(r)$, and (c) junction-terminal pairs, $g^{\,}_{\rm JT}(r)$, evaluated at different times during the relaxation of the fingerprint labyrinthine pattern. The development and sharpening of short-range peaks reflect the emergence of strong local correlations between defects, while the absence of long-range structure indicates the lack of global positional order even at late times.}
\label{fig:g_r}
\end{figure}

To quantify the spatial organization of defects during coarsening, we analyze the pair distribution function (PDF) $g_{\alpha\beta}(r)$. A closely related approach has recently been employed to characterize morphological transitions in labyrinthine patterns in magnetic thin films~\cite{shimizu2024}. Originally developed in the study of simple liquids and later extended to a broad range of systems—including crystalline solids, glasses, and colloidal suspensions—the PDF provides a robust statistical measure of local structural correlations~\cite{allen2017,hansen1990,rapaport2004}. In the present context, $g_{\alpha\beta}(r)$ quantifies the normalized probability of finding a defect of type $\beta$ at a distance $r$ from a reference defect of type $\alpha$, relative to that expected for a spatially uniform distribution at the same defect density. In ordered systems, pronounced peaks in $g(r)$ signal well-defined characteristic length scales, whereas in disordered systems they reveal short-range correlations and emergent local motifs. Extending this framework to ensembles of point-like defects in pattern-forming systems thus enables a quantitative characterization of defect interactions that goes beyond qualitative real-space snapshots. Formally, we define
\begin{align}
	\label{eq:g_r}
	g_{\alpha \beta}(r) = \frac{1}{\rho_\beta N_\alpha} \Bigg\langle \sum_{i=1}^{N_\alpha}\sum_{j=1}^{N_\beta} \delta\big(r - |\mathbf{r}_i^\alpha - \mathbf{r}_j^\beta|\big) \Bigg\rangle,
\end{align}
where $N\alpha$ and $N\beta$ denote the numbers of defects of type $\alpha$ and $\beta$, respectively, $\rho_\beta$ is the average density of $\beta$ defects, and $\langle\cdots\rangle$ indicates an ensemble or time average. This normalization ensures that $g_{\alpha\beta}(r)\to 1$ at large separations, where positional correlations are lost and the defects become effectively uncorrelated.

\begin{figure}
\centering
\includegraphics[width=0.99\columnwidth]{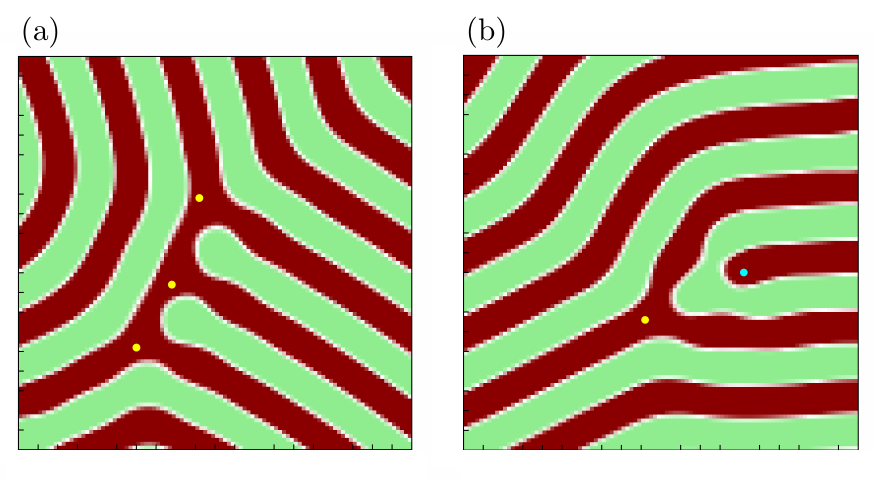}
\caption{Metastable local defect configurations in the fingerprint labyrinthine pattern. (a) Stable clusters and short arrays of junctions pinned near regions of stripe frustration, where competing stripe orientations inhibit further coarsening. (b) A bound junction-terminal pair forming a robust local configuration. Both structures represent long-lived metastable states that contribute to the arrested relaxation and residual defect density at late times.}
\label{fig:meta-stable-defects}
\end{figure}

Applying this analysis to our simulations reveals clear and systematic statistical signatures of defect interactions. Figure~\ref{fig:g_r} presents the PDFs for different combinations of defect species, evaluated at several representative times during the relaxation process. For all three cases; junction–junction (J–J), terminal–terminal (T–T), and junction–terminal (J–T), the PDFs exhibit pronounced short-range correlations, manifested as a sequence of well-defined peaks separated by minima at increasing distances $r$. At larger separations ($r \gtrsim 50$ units), all pair distribution functions gradually approach unity, signaling the loss of positional correlations beyond the intrinsic stripe wavelength. This behavior stands in sharp contrast to the coarsening of topological defects such as vortices in the XY model, where weakly interacting and largely uncorrelated vortices give rise to an almost featureless PDF, $g(r)\approx 1$, beyond a short-distance cutoff.

Junction–junction correlations display a pronounced first peak in $g^{\,}_{\rm JJ}(r)$ at $r \approx 20$ spatial units, indicating a preferred separation associated with tightly bound junction clusters. These clusters typically consist of pairs or triplets of junctions that become pinned near frustrated regions of the stripe network, as illustrated in Fig.~\ref{fig:meta-stable-defects}(a). Terminal–terminal correlations, quantified by $g^{\,}_{\rm TT}(r)$, exhibit a similarly well-defined first peak at comparable separations. This feature reflects the tendency of terminals to form correlated arrangements mediated by the surrounding stripe geometry, rather than appearing as isolated and independently distributed defects.

By contrast, junction–terminal correlations peak at a slightly larger separation, $r \approx 25$ units, consistent with the formation of topologically neutral dipoles composed of $-1/2$ and $+1/2$ defects. An example of such a J–T dipole consisting of a junction (topological charge $-1/2$) bound to a terminal (charge $+1/2$), is shown in Fig.~\ref{fig:meta-stable-defects}(b). These bound states experience an effective attractive interaction and are frequently observed to undergo mutual annihilation at later stages of the dynamics, thereby directly contributing to the reduction of defect density during coarsening.

As coarsening proceeds, the gradual growth of the first few peaks of $g(r)$ signals increasingly strong short-range correlations among the remaining defects. Notably, the enhancement of the first peak is more pronounced than that of the second, which in turn grows faster than the third, and so on. This hierarchical amplification of short-range structure reflects the progressive reduction in the size and complexity of defect clusters through successive annihilation and rearrangement processes, as illustrated in Fig.~\ref{fig:anni}. Concurrently, coarsening manifests itself in the slow separation and dilution of defect clusters such as the extended junction array shown in Fig.~\ref{fig:meta-stable-defects}(a), as the characteristic length scale of the fingerprint labyrinthine pattern continues to increase with time.

Overall, the PDF analysis demonstrates that the coarsening dynamics of fingerprint labyrinthine patterns are governed by strongly correlated, short-range defect interactions rather than by weakly interacting, freely diffusing defects. This statistical characterization thus provides a compact and quantitative bridge between the microscopic organization of point defects and the emergent growth of length scales during pattern coarsening.

\section{Summary and outlook}

\label{sec:summary}

In this work, we have presented a detailed study of the coarsening dynamics of fingerprint labyrinthine patterns generated by the Turing-Swift-Hohenberg equation. By combining conventional Fourier-based diagnostics with real-space, defect-resolved analysis, we have shown that the late-stage relaxation of fingerprint labyrinths differs qualitatively from both conventional symmetry-breaking phases and simple stripe patterns. While the global structure factor retains a persistent ring signature reflecting robust local wavelength selection, the narrowing of this ring proceeds slowly and eventually saturates, signaling arrested or strongly constrained coarsening rather than the emergence of long-range orientational order.

A central result of this study is the identification of localized junction and terminal defects as the primary agents controlling coarsening in fingerprint labyrinths. Using a template-matching convolutional neural network (TM-CNN), we were able to systematically detect, classify, and track these point-like defects throughout the relaxation process. Analysis of defect densities, interconversion events, and pair distribution functions reveals that defect motion is highly constrained by the surrounding stripe geometry and proceeds through localized stripe breaking and reconnection rather than diffusive random walks. The resulting dynamics lead to slow, algebraic relaxation toward a frozen state with a finite residual defect density, providing a microscopic explanation for the breakdown of conventional coarsening laws in this regime.

More broadly, fingerprint labyrinthine patterns represent an intermediate and physically transparent regime within a broader hierarchy of labyrinthine states. As shown in prior work on labyrinthine pattern transitions, increasing the control parameter leads from fingerprint-type labyrinths to more strongly disordered glassy labyrinths, characterized by nearly frozen defect configurations and eventually to scurfy labyrinths, where spot-like structures proliferate within the patterned background. Understanding coarsening in fingerprint labyrinths thus provides a natural stepping stone toward addressing the far more constrained and heterogeneous dynamics of these more complex labyrinthine phases, where defect motion is even more strongly frustrated or effectively arrested. Beyond the TSH equation, similar labyrinthine morphologies arise in a wide range of driven, dissipative systems, including reaction–diffusion media, magnetic films, liquid crystals, and biological patterning. Extending defect-based and real-space approaches to these systems offers a promising route toward a unified description of phase-ordering dynamics in pattern-forming systems that lie outside the conventional symmetry-breaking paradigm.

Finally, this study highlights the growing role of machine-learning assisted approaches in the quantitative analysis of phase-ordering dynamics in complex pattern-forming systems. The TM-CNN framework provides a powerful and efficient means of extracting defect-level information that is difficult or impossible to obtain using global spectral measures alone. More generally, combining physics-informed diagnostics with modern machine-learning tools, such as convolutional networks, graph-based representations of defect networks, and unsupervised clustering methods, opens new avenues for identifying emergent structures, dynamical bottlenecks, and hidden length scales in non-equilibrium systems. We expect such hybrid approaches to play an increasingly important role in extending phase-ordering theory to regimes characterized by geometric frustration, complex topology, and slow, non-diffusive dynamics.

\begin{acknowledgments}
The work of S.G. and G.W.C. was partially supported by the U.S. Department of Energy, Office of Basic Energy Sciences, under Contract No. DE-SC0020330. The work of V.Y.O. and H.Y.K. was partially supported by the São Paulo Research Foundation (FAPESP), Brazil (Grants No. 2024/10263-3 and No. 2025/03683-9), and by the National Council for Scientific and Technological Development (CNPq), Brazil (Grant No. 300724/2025-0). The work of B.S.S. was partially supported by the National Science Foundation under Grant No. DMR-2016909. S.G. also acknowledges the support of Research Computing at the University of Virginia.
\end{acknowledgments}

\bibliography{ref}

\end{document}